\documentclass[aps,prd,twocolumn,showpacs,preprintnumbers,nofootinbib,amsmath]{revtex4}

\usepackage[english]{babel}
\usepackage{amsmath,amssymb}
\usepackage{graphicx}
\usepackage[sort&compress]{natbib}
\usepackage{slashed}            
\allowdisplaybreaks

\bibpunct{[}{]}{,}{n}{}{,}

\newcommand{\bra}[1]{\ensuremath{\langle #1 |}}   
\newcommand{\ket}[1]{\ensuremath{| #1 \rangle}}   

\def\be{\begin{equation}}
\def\ee{\end{equation}}
\def\bea{\begin{eqnarray}}
\def\eea{\end{eqnarray}}

\renewcommand{\vec}[1]{{\mathbf{#1}}}

\usepackage{xcolor}
\definecolor{red}{rgb}{1.0, 0, 0}

\begin{document}
\def\cM{\mathcal{M}}
\def\prop#1{{\cal P}_{#1}}
\def\sutqm{(s_{13}+s_{14})}
\def\sdtqm{(s_{23}+s_{24})}
\def\sucsm{(s_{15}+s_{16})}
\def\sdcsm{(s_{25}+s_{26})}
\def\sutq{{s}_{134}}
\def\sdtq{{s}_{234}}
\def\sucs{{s}_{156}}
\def\sdcs{{s}_{256}}
\def\sud{{s}_{12}}
\def\stq{{s}_{34}}
\def\scs{{s}_{56}}
\def\sut{{s}_{13}}
\def\suq{{s}_{14}}
\def\sdt{{s}_{23}}
\def\sdq{{s}_{24}}
\def\suc{{s}_{15}}
\def\sus{{s}_{16}}
\def\sdc{{s}_{25}}
\def\sds{{s}_{26}}
\def\duxtqxd{1}
\def\ddxuxtq{2}
\def\duxdxtq{3}
\def\cuxd{1}
\def\cudxtq{2}
\def\cuxtq{3}
\def\cdxtq{4}
\def\cuxcs{5}
\def\cdxcs{6}
\def\cD{\cal D}
\def\cP{\cal P}
\def\cg{c_\Gamma}
\def\gW{g_W}
\def\rG{r_\Gamma}
\def\jb{i_{b}}
\def\jt{i_{t}}
\def\tb{\bar{t}}
\def\eb{\bar{e}}
\def\mub{\bar{\mu}}
\def\qb{\bar{q}}
\def\spaa#1.#2.#3{\langle\mskip-1mu{#1}|#2|{#3}\mskip-1mu\rangle}
\def\spbb#1.#2.#3{[\mskip-1mu{#1}|#2|{#3}\mskip-1mu]}
\def\spa#1.#2{\left\langle#1\,#2\right\rangle}
\def\spb#1.#2{\left[#1\,#2\right]}
\def\spab#1.#2.#3{\left\langle#1|#2|#3\right]}
\def\spba#1.#2.#3{\left[#1|#2|#3\right\rangle}
\def\spbab#1.#2.#3.#4{[\mskip-1mu{#1}
                  | #2  #3 | {#4}\mskip-1mu]}
\def\spaba#1.#2.#3.#4{\langle\mskip-1mu{#1}
                  | #2  #3 | {#4}\mskip-1mu\rangle}
\def\Wzb{\bar{W}_0}
\def\Pzb{\bar{P}_0}
\def\Pthreeb{\bar{P}_3}
\def\P3b{\bar{P}_3}
\def\Ypb{\bar{Y}_p}
\def\Ypz{{Y}_p(z)}
\def\Ywb{\bar{Y}_w}
\def\Ppb{\bar{P}_+}
\def\Pmb{\bar{P}_-}
\def\Wpb{\bar{W}_+}
\def\Wmb{\bar{W}_-}
\def\cO{{\cal O}}
\def\li{{\rm Li_2}}
\def\g0{\gamma_0}
\def\gp{\gamma^{+}}
\def\gm{\gamma^{-}}
\def\lp{\gamma^{+}}
\def\lm{\gamma^{-}}
\def\xp{x_{+}}
\def\xm{x_{-}}
\def\bentarrow{\:\raisebox{1.3ex}{\rlap{$\vert$}}\!\rightarrow}                 
\def\dkp#1#2#3#4{
\begin{array}{r c l}
#1 & \rightarrow & #2#3 \\
 & & \phantom{\; #2}\bentarrow #4
\end{array}}                                                                    
\def\bothdk#1#2#3#4#5{
\begin{array}{r c l}                                                            
#1 & \rightarrow & #2#3 \\
 & & \:\raisebox{1.3ex}{\rlap{$\vert$}}\raisebox{-0.5ex}{$\vert$} 
\phantom{#2}\!\bentarrow #4 \\
 & & \bentarrow #5                                                              
\end{array}                                                                     
}                                                                               
\newcommand{\kirill}{\colour{red}}
\newcommand{\comment}[1]{{\bf [#1]}}
\newcommand{\beq}{\begin{equation}}
\newcommand{\eeq}{\end{equation}}
\newcommand{\beqn}{\begin{eqnarray}}
\newcommand{\eeqn}{\end{eqnarray}}
\newcommand{\bi}[1]{\bibitem{#1}}
\newcommand{\fr}[2]{\frac{#1}{#2}}
\newcommand{\non}{\nonumber}
\newcommand{\nn}{\nonumber}
\newcommand{\Et}{E_t}
\newcommand{\Pt}{P_t}
\newcommand{\pt}{p_t}
\newcommand{\pb}{p_b}
\newcommand{\pw}{p_W}
\newcommand{\pg}{p_g}
\newcommand\tpW        {{\tilde p}_W}
\newcommand\tpb        {\tilde p_b}
\newcommand{\ar}{\mbox{$\rightarrow$}}
\def\ra{\rightarrow}

\newcommand{\slsh}{\rlap{$\;\!\!\not$}}     
\def\amuh{a_\mu^{{\mathrm had}}}
\def\vec#1{{\mbox{\boldmath$#1$}}}
\def\ket#1{\vert #1 \rangle}
\def\bra#1{\langle #1 \vert}
\newcommand{\as}{\alpha_S}
\newcommand{\p}{\mbox{$\vec{p}$}}
\newcommand{\q}{\mbox{$\vec{q}$}}
\newcommand{\pp}{\mbox{$\vec{p}'$}}
\newcommand{\rp}{\mbox{$\vec{r}'$}}
\newcommand{\kp}{\mbox{$\vec{k}'$}}
\newcommand{\e}{\mbox{$\vec{e}$}}
\newcommand{\s}{\mbox{$\vec{s}$}}
\newcommand{\Li}{{\rm Li}}
\newcommand{\lsim}{\mbox{\raisebox{-0.3ex}{%
\footnotesize $\:\stackrel{<}{\sim}\:$}} }
\newcommand{\gsim}{\mbox{\raisebox{-0.3ex}{%
\footnotesize $\:\stackrel{>}{\sim}\:$}} }
\newcommand{\lb}{\left (}
\newcommand{\rb}{\right )}
\newcommand{\ep}{\epsilon}
\newcommand{\vep}{\epsilon}
\newcommand{\dd}{{\rm d}}
\newcommand{\om}{\omega}

\newcommand{\sS}{\mbox{$\vec{\sigma}\vec{\sigma}'$}}
\newcommand{\si}{\mbox{$\vec{\sigma}$}}
\newcommand{\vgamma}{\mbox{$\vec{\gamma}$}}
\newcommand{\vxi}{\mbox{$\vec{\xi}$}}

\newcommand{\pop}[1]{\mbox{$\Lambda_+(#1)$}}
\newcommand{\nep}[1]{\mbox{$\Lambda_-(#1)$}}
\newcommand{\Dafne}{DA$\Phi$NE}
\newcommand{\mar}{\marginpar{***}}
\def\spab#1.#2.#3{\langle\mskip-1mu{#1}
                  | #2 | {#3}\mskip-1mu]}
\def\spba#1.#2.#3{[\mskip-1mu{#1}
                  | #2 | {#3}\mskip-1mu\rangle}
\def\spa#1.#2{\langle#1\,#2\rangle}
\def\spb#1.#2{[#1\,#2]}

\def\dk#1#2#3{
\begin{array}{r c l}
#1 & \rightarrow & #2 \\
 & & \bentarrow #3
\end{array}
}

\title{Bounding the Higgs width at the LHC: complementary results from $H \to WW$}

\author{John M. Campbell}
\email[Email: ]{johnmc@fnal.gov}
\affiliation{Fermilab, Batavia, IL 60510, USA}
\author{R. Keith Ellis}
\email[Email: ]{ellis@fnal.gov}
\affiliation{Fermilab, Batavia, IL 60510, USA}
\author{Ciaran Williams}
\email[Email: ]{ciaran@nbi.dk}
\affiliation{Niels Bohr International Academy and Discovery Center,
The Niels Bohr Institute, Blegdamsvej 17, DK-2100 Copenhagen \O, Denmark}
\preprint{FERMILAB-PUB-13-553-T}

\begin{abstract}
We investigate the potential of the process $gg \to H \to WW$ to provide bounds on the Higgs width. 
Recent studies using off-shell $H\rightarrow ZZ$ events have shown that Run 1 LHC data
can constrain the Higgs width, $\Gamma_H < (25-45) \, \Gamma_{H}^{\rm SM}$.  
Using 20 fb$^{-1}$ of 8 TeV ATLAS data, we estimate a bound on the Higgs boson width from the
$WW$ channel between $\Gamma_H < (100-500) \, \Gamma_{H}^{\rm SM}$.  
The large spread in limits is due to the range of
cuts applied in the existing experimental analysis. The stricter cuts designed to search for the on-shell 
Higgs boson limit the potential number of off-shell events, weakening the constraints.
As some of the cuts are lifted the bounds improve. We show that there is potential in the high transverse
mass region to produce upper bounds of the order of $(25-50) \, \Gamma_{H}^{\rm SM}$, depending strongly on the level
of systematic uncertainty that can be obtained.
Thus, if these systematics can be controlled, a constraint on the Higgs boson width
from the $H \to WW$ decay mode can complement a corresponding limit from $H \to ZZ$. 
\end{abstract}

\maketitle

\section{Introduction}

The discovery of a boson~\cite{Aad:2012tfa,Chatrchyan:2012ufa} 
which broadly agrees with the predictions of a SM 126 GeV Higgs boson~\cite{ATLAS:2013mma,Chatrchyan:2012jja,Aad:2013xqa,Aad:2013wqa,CMS:yva},
represents a tremendous achievement for the LHC experiments, and the theoretical predictions of the Standard Model (SM). The continued 
study of the Higgs boson will provide a detailed understanding of its couplings to the other SM particles. Extraction of these parameters 
is complicated by the form of the cross section in the narrow width approximation (NWA), 
\begin{eqnarray}
\sigma_{i\rightarrow H \rightarrow f} \sim \frac{g_{i}^2g_{f}^2}{\Gamma_H}.
\label{eq:NWAxs}
\end{eqnarray} 
In this approximation the cross section is invariant under the simultaneous rescaling $g_x \rightarrow \xi g_x$
and $\Gamma_H\rightarrow \xi^4 \Gamma_H$. Therefore attempts to extract coupling information from individual cross section
measurements require an assumption of the width, or its direct measurement. 

Due to the expected scale of the Higgs width ($4$~MeV), it is hard to extract its value 
directly at the LHC because of the inherent detector resolution scale ($\sim$  $1$--$2$~GeV).
Ultimately a precision measurement may be provided by a lepton collider,
either by measurement of the invisible Higgs branching ratio ($e^+e^-$)~\cite{Han:2013kya} or a direct threshold scan ($\mu^+\mu^-$)~\cite{Han:2012rb,Conway:2013lca}. 
Until such a time the LHC
can follow a number of alternative strategies that provide less direct constraints. One possibility is to combine experimental results 
across all Higgs boson production and decay channels~\cite{Dobrescu:2012td}.   This provides rather stringent limits,
$\Gamma_{H} \lesssim (3-4) \, \Gamma_{H}^{\rm SM}$, albeit with the caveat of mild theoretical assumptions.  An alternative approach~\cite{Dixon:2013haa}
is to use the mass shift induced by the interference between the Higgs signal 
and the SM continuum in $H\rightarrow \gamma\gamma$~\cite{Dixon:2003yb,Martin:2012xc,Martin:2013ula}. 
By measuring the mass shift relative to 
a second channel, one can constrain the couplings and thus the total Higgs width.
If the relative mass difference could be measured to $\mathcal{O}(50-100)$~MeV with $3~{\rm ab}^{-1}$ of total LHC data
then the upper bound on the width would be $\Gamma_{H} \lesssim (10-20) \, \Gamma_{H}^{\rm SM}$~\cite{Dixon:2013haa}.

In a recent paper~\cite{Caola:2013yja}
Caola and Melnikov proposed a further mechanism to constrain the Higgs width using the $H \to ZZ \to 4\ell$ decay.  
The method relies on the fundamental difference between the on-shell cross section shown in Eq.~(\ref{eq:NWAxs}), and 
the off-shell cross section. Away from the resonance region the Higgs propagator is dominated by the $(s_H-m_H^2)$ 
term and therefore the number of off-shell Higgs events depends on the coupling rescaling factor $\xi$. This allows 
a determination of upper limits on $\xi$ and hence the Higgs boson width.
With existing LHC data this method can constrain the Higgs 
width at the level of $25\, \Gamma_H$.  Potential improvements from kinematic discriminants should sharpen the 
constraints to around $15\, \Gamma_H$~\cite{Campbell:2013una}.

Given its importance it is natural to investigate other channels that may be amenable to a similar analysis.
At first glance the lack of mass resolution in the channel $H\rightarrow WW \rightarrow 2\nu 2\ell$
may appear to undermine the method, which relies on a separation between the peak and off-shell regions. 
However, the crux of the method relies only on the existence of this separation and not on its exact reconstruction.
Any variable that has well-defined ``on'' and ``off'' peak regions can be sensitive to the rescaling parameter $\xi$.
One such variable is the transverse mass $M_T$~\cite{ATLAS:2013wla}  
defined through,
\begin{eqnarray}
M_T^2 = ({E_T^{miss}}+{E_T^{\ell\ell}})^2+ |{\bf{p}}_T^{\ell\ell}+{\bf{E}}_T^{miss}|^2 
\label{eq:MTdef}
\end{eqnarray}
where $E_T^{\ell\ell}=(|{\bf{p}}_T^{\ell\ell}|^2+m_{\ell\ell}^2)^{1/2}$. We note that at LO in perturbation theory (for $WW$ final states) 
the second term in $M_T$ is identically zero. As such the on-shell events necessarily
have $M_T < m_H$, resulting in a prominent edge in the distribution. For the off-shell events
no such restriction applies. Therefore a comparison of the cross sections in the
two regions $M_T < m_H$ and $M_T > m_H$ provides information on the Higgs boson width. 
  
The $WW$ channel has several advantages when compared to the $ZZ$ decay.  Firstly, the threshold for
producing two real $W$ bosons is much closer than the threshold for two $Z$ bosons. Secondly the branching ratio
into the leptonic final state is much larger.  Combined, this means that the number of Higgs events is about two orders of
magnitude larger since
${\rm Br}(H \to WW) \times {\rm Br}(W \to \ell\nu)^2 = 2.7 \times 10^{-3}$ while
${\rm Br}(H \to ZZ) \times {\rm Br}(Z \to \ell^+\ell^-)^2 = 3.2 \times 10^{-5}$.
On the other hand this channel has several disadvantages.  The primary concern is that it suffers from a more extensive
list of backgrounds that are both larger and more complicated to remove than in the $ZZ$ analysis.
The top background is overwhelming in the inclusive sample but 
a jet veto can be used to isolate Higgs bosons produced through gluon fusion.  Using the 7 and 8 TeV 
data ATLAS~\cite{ATLAS:2013wla} and CMS~\cite{CMS:wwa}  have found evidence for a $126$~GeV Higgs boson in this channel with
a significance of $3.8$ and $3.1$ standard deviations respectively. Very recently CMS have updated their results 
to include the full 8 TeV data set, improving the significance to 4.3 standard deviations~\cite{Chatrchyan:2013iaa}.
     
In this paper we apply the technique of ref.~\cite{Caola:2013yja} to the $WW$ channel, and investigate its 
potential to bound the Higgs width at the LHC.  In section~\ref{sec:Results} we briefly
describe our calculations and present our results for the off-shell cross section as a function of 
the Higgs width rescaling. Section~\ref{sec:ATLAS} uses recent results from ATLAS to bound the Higgs width and discusses
future improvements that could be made to facilitate a stronger bound. Finally we draw our conclusions in section~\ref{sec:conc}.

\section{Results}
\label{sec:Results}

In order to correctly model the off-shell spectrum it is crucial to include the effects of the interference 
between the SM continuum and the Higgs production of $WW$. The SM continuum proceeds through a 
gluon-induced loop of fermions. This process has a rich history, with the first calculations performed 
(for on-shell $W$ bosons) in the late 1980's~\cite{Dicus:1987dj,Glover:1988fe}. Off-shell effects, including 
vector boson decays were presented in~\cite{Binoth:2005ua} and updated to include the full mass of the
top and bottom quarks in ~\cite{Binoth:2006mf}. Full analytic results for helicity amplitudes with massless
quarks were presented in ref.~\cite{Campbell:2011bn}, which made use of the $e^+e^-\to 4$ parton amplitudes of 
ref. \cite{Bern:1997sc}. These results were extended to include the mass of the top quark in~\cite{Campbell:2011cu} 
and a detailed study of the effect on the Higgs interference (for searches over a wide range of Higgs masses) was 
presented. More recently the interference has been studied in the context of a $125$~GeV Higgs boson~\cite{Kauer:2012hd,Kauer:2013qba}.
Higher order corrections to the interference,
computed using a soft-collinear approximation to NLO and NNLO, have been investigated in ref~\cite{Bonvini:2013jha}.
 
\begin{table}
\begin{tabular}{|c|c|c|c|}
\hline
$m_H$               & 126 GeV     & $\Gamma_H$ & 0.004307 GeV \\
$m_W$               & 80.398 GeV  & $\Gamma_W$ & 2.1054 GeV \\
$m_t$               & 173.2 GeV   & $m_b$      & 4.75 GeV \\
$e^2$               & 0.0949563   & $g_W^2$    & 0.4264904 \\ 
$\sin^2\theta_W$    & 0.2226459   & $G_F$      &$0.116639\times10^{-4}$ \\
\hline
\end{tabular}
\caption{Mass, width and electroweak parameters used to produce the results in this paper.
\label{parameters}}
\end{table}
In this paper we use the implementation of $\nu_e e^+ \mu^- \bar\nu_\mu$ production in the parton level integrator MCFM 6.7.
This includes the effect of massive top and bottom quarks in the Higgs amplitude, while the continuum
amplitude accounts for the effect of the top quark mass while leaving the $b$-quarks massless.
Our electroweak parameters are listed in Table~\ref{parameters} and correspond to the default choices
in MCFM.

\begin{center}
\begin{table}
\begin{tabular}{|l@{\hskip 0.6cm}l|}
\hline
 \multicolumn{2}{|l|}{$|\eta_{e}| < 2.47$ excluding $1.37 < |\eta_e| < 1.52$}  \\  
 $|\eta_{\mu}| < 2.5 $                               & $10$~GeV$~< m_{\ell\ell} < 50$~GeV                              \\
 $p_T^{\ell} {\rm (hardest)}> 25 $~GeV               & $\Delta \phi_{\ell\ell}<1.8$                                  \\
 $p_T^{\ell} {\rm (softest)}> 15 $~GeV               & $E_{T,miss}^{rel}> 25$~GeV                                      \\
 $p_T^{\ell\ell}>30$~GeV                             & $|\Delta \phi_{\ell\ell, MET}|> \pi/2$                          \\ 
\hline
\end{tabular}
\caption{The cuts used in this paper, referred to as ``full'' cuts, designed to mimic the $e\mu+\mu e$ analysis of ATLAS~\cite{ATLAS:2013wla}.}
\label{ATLAScuts}
\end{table}
\end{center}
We will present results for a set of cuts designed to mimic the analysis reported by
the ATLAS collaboration in Ref.~\cite{ATLAS:2013wla}.
A list of the cuts that we apply is given in Table~\ref{ATLAScuts} (``full'' cuts).
The number of off-shell Higgs events is particularly sensitive to two of these cuts, namely those on the dilepton invariant mass $m_{\ell\ell}$, 
and the azimuthal angle $\phi_{\ell\ell}$. These are subject to the upper bounds,
\begin{equation}
m_{\ell\ell} < 50 \,{\rm{GeV}} \,, \quad \Delta \phi_{\ell\ell} < 1.8 \,.
\label{eq:mphicuts}
\end{equation}
In order to regain sensitivity to the off-shell region we will also consider the scenario
in which the above dilepton invariant mass cut is removed and the case in which both of these cuts are removed (``basic'' cuts).

\begin{center} 
\begin{figure*}
\includegraphics[scale=0.6,angle=0]{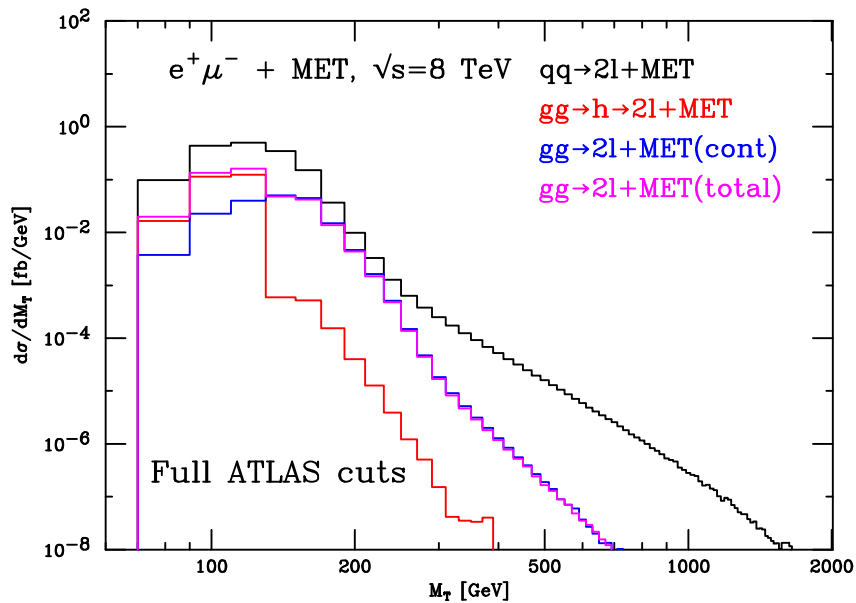} 
\caption{Overall picture at 8 TeV (colour online), with the full
ATLAS cuts described in the text imposed.}
\label{Bigpicture8full}
\end{figure*}
\end{center}
\begin{center} 
\begin{figure*}
\includegraphics[scale=0.6,angle=0]{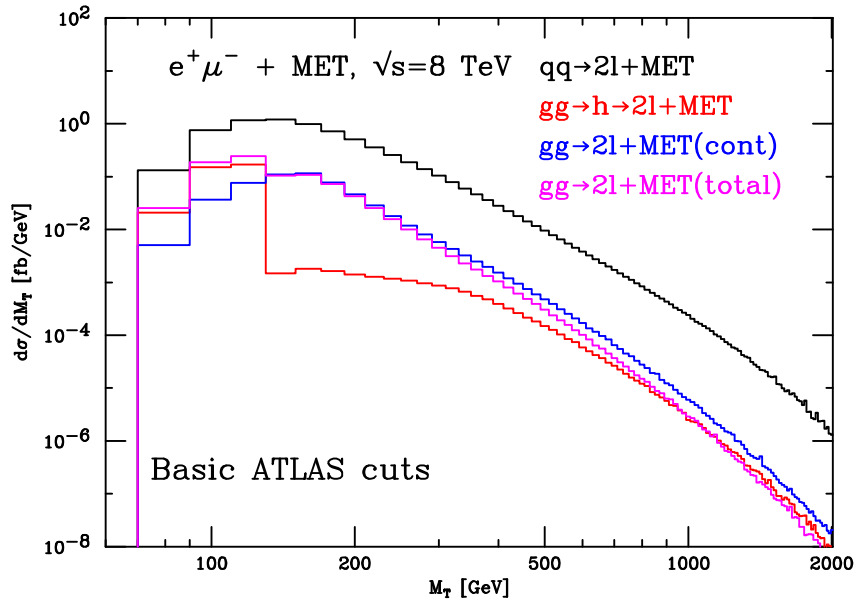} 
\caption{Overall picture at 8 TeV (colour online), under the 
basic ATLAS cuts, i.e. when neither the $m_{\ell\ell}$
nor the $\Delta\phi_{\ell\ell}$ cut in Eq.~(\ref{eq:mphicuts}) has been applied.}
\label{Bigpicture8basic}
\end{figure*}
\end{center}
\begin{figure*}
\begin{center} 
\includegraphics[scale=0.6,angle=0]{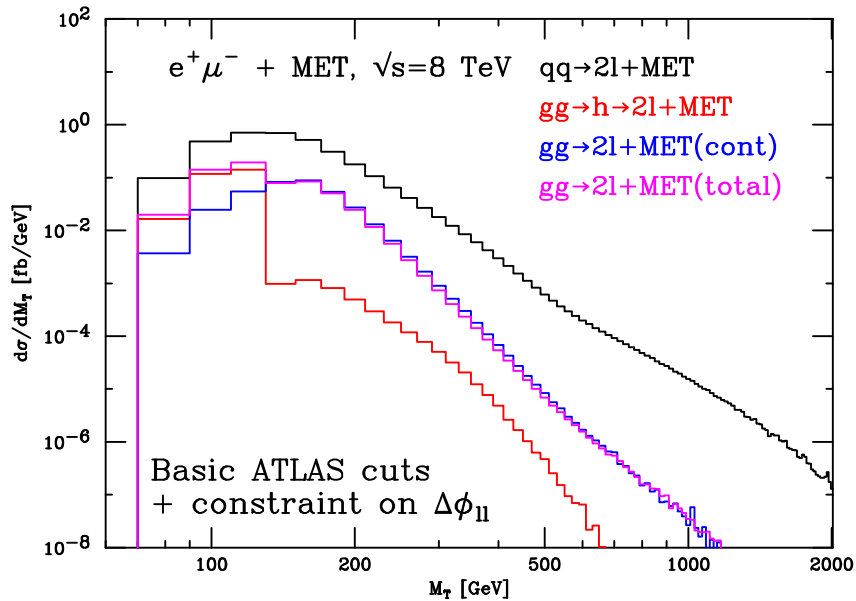} 
\caption{Overall picture at 8 TeV (colour online). In this figure the ATLAS cuts described in the text
have been imposed, including the $\Delta\phi_{\ell\ell}$ cut in Eq.~(\ref{eq:mphicuts}) but removing the
constraint on the maximum invariant mass of the dilepton pair.}
\label{Bigpicture8nomll}
\end{center}
\end{figure*}

We present the contribution of the Higgs signal and gluon-gluon box diagrams, together with the dominant
$q\bar q \to WW \to 2\ell2\nu$ background process, in Figs.~\ref{Bigpicture8full} and~\ref{Bigpicture8basic}.
Results are shown as a function of the transverse mass $M_T$ defined in Eq.~(\ref{eq:MTdef}), for $m_H=126$~GeV at $\sqrt{s}=8$~TeV and for
the decay of the $W^+$ into an electron and $W^-$ into a muon.
The renormalization and factorization scales are set equal to $\hat s/2$, where $\sqrt{\hat s}$ is the partonic center-of-mass energy.
Note that, in our calculations, $\hat s$ is identical to the four-momentum squared of the final state.
Figs.~\ref{Bigpicture8full} and~\ref{Bigpicture8basic} contain a mixture of orders in perturbation
theory. The $q \bar{q}$ process is included at lowest order in perturbation theory $O(g_W^8)$,
whereas the other processes are included at $O(g_W^8 g_s^4)$, i.e.\ they  are next-to-next-to 
leading with respect to the $q\bar{q}$ process, but enhanced by large gluon fluxes at the LHC. 
The kinematic edge at the Higgs boson mass is visible.
At high values of $M_T$, and hence of $\sqrt{\hat s}$, the effect of
the interference is destructive and cancels the leading high energy behaviour of the
$gg \to W^+W^- \to \nu_e e^+ \mu^- \bar\nu_\mu$ process.
Comparing Figs.~\ref{Bigpicture8full} and~\ref{Bigpicture8basic} it is clear that the full ATLAS cuts greatly
reduce the impact of the off-peak region.  We note that one
way of mitigating this suppression in the future might be to apply only the $\Delta\phi_{\ell\ell}$ cut in Eq.~(\ref{eq:mphicuts}),
removing the upper bound on the dilepton invariant mass.  As can be seen from Fig.~\ref{Bigpicture8nomll}, this has the advantage of
maintaining a strong rejection of the continuum background while accepting more of the high-$M_T$ tail.

Table~\ref{xsec8tabledyn} shows the cross section in bins of transverse mass, corresponding to
the Higgs peak region $M_T<130$~GeV and in two off-shell regions defined by $M_T>130$~GeV
and $M_T>300$~GeV. In this table the cross sections correspond to the following matrix elements,
\beq
\sigma^H: |\cM_H|^2 \,, \quad \sigma^I: |\cM_H+\cM_C|^2 - |\cM_C|^2 - |\cM_H|^2  \,,
\label{eq:sigmadefn}
\eeq
where $\cM_H$ is the Higgs production amplitude and $\cM_C$ is the amplitude for the continuum background.
Assuming the Standard Model width for the Higgs boson, the effect of
the Higgs diagrams is dominated by the interference contribution and the cross section in the off-peak region $M_T>130$~GeV
is reduced.  Under the full set of ATLAS cuts this effect is very small, at the level of $3$\% of the cross section in the peak
region.  The effect becomes larger as the $m_{\ell\ell}$ and $\Delta\phi_{\ell\ell}$ cuts are removed, resulting in a
sizeable $11$\% effect in the case of the basic cuts.
\begin{table*}
\begin{tabular}{|c||c|c||c|c||c|c|}
\hline
 &\multicolumn{2}{c||}{\parbox[c]{0.2\linewidth}{$M_T <130$~GeV}} &  
\multicolumn{2}{c||}{\parbox[c]{0.2\linewidth}{$M_T>130$~GeV}} &
\multicolumn{2}{c|}{\parbox[c]{0.2\linewidth}{$M_T>300$~GeV}} \\
Cuts &   \parbox[c]{0.1\linewidth}{$\sigma^H$}
& \parbox[c]{0.1\linewidth}{$\sigma^{I}$}
& \parbox[c]{0.1\linewidth}{$\sigma^H$}
& \parbox[c]{0.1\linewidth}{$\sigma^{I}$}
& \parbox[c]{0.1\linewidth}{$\sigma^H$}
& \parbox[c]{0.1\linewidth}{$\sigma^{I}$} \\
\hline
\hline
full                              & 5.06 & -0.0778 & 0.0262 & -0.173 & -      & - \\
\hline
basic  + $\Delta\phi_{\ell\ell}$  & 5.52 & -0.0924 & 0.0844 & -0.483 & 0.0021 & -0.00888 \\
\hline
basic                             & 6.85 & -0.117  & 0.328  & -1.07  & 0.104  & -0.240 \\
\hline
\hline
\end{tabular}
\caption{Fiducial cross sections for $pp \to W^+W^- \to \nu_e e^+ \mu^- \bar\nu_\mu$ in fb,
for the LHC operating at $\sqrt s=8$~TeV.
All cross-sections are computed with
leading order MSTW 2008 parton distribution functions~\cite{Martin:2009iq} and
renormalization and factorization scales set equal to $\hat s /2$.}
\label{xsec8tabledyn}
\end{table*}

In Table~\ref{xsec13tabledyn} we show the expected cross sections, under the same set of cuts, at $\sqrt{s}=13$~TeV.
At this higher operating energy the value of the peak cross section approximately doubles while the off-peak region
grows by a factor of three or more.  This is due to the growth of the gluon pdf at small $x$, to which the tails of the
$M_T$ distribution are more sensitive.  The relative enhancement of the off-peak region suggests that the types of
analyses that we describe in this paper will be more effective at $\sqrt{s}=13$~TeV.
\begin{table*}
\begin{tabular}{|c||c|c||c|c||c|c|}
\hline
 &\multicolumn{2}{c||}{\parbox[c]{0.2\linewidth}{$M_T <130$~GeV}} &  
\multicolumn{2}{c||}{\parbox[c]{0.2\linewidth}{$M_T>130$~GeV}} &
\multicolumn{2}{c|}{\parbox[c]{0.2\linewidth}{$M_T>300$~GeV}} \\
Cuts &   \parbox[c]{0.1\linewidth}{$\sigma^H$}
& \parbox[c]{0.1\linewidth}{$\sigma^{I}$}
& \parbox[c]{0.1\linewidth}{$\sigma^H$}
& \parbox[c]{0.1\linewidth}{$\sigma^{I}$}
& \parbox[c]{0.1\linewidth}{$\sigma^H$}
& \parbox[c]{0.1\linewidth}{$\sigma^{I}$} \\
\hline
\hline
full                              & 11.3 & -0.195 & 0.0658 & -0.431 & -       & -0.000185 \\
\hline
basic  + $\Delta\phi_{\ell\ell}$  & 12.3 & -0.233 & 0.222  & -1.25  & 0.00698 & -0.0283 \\
\hline 
basic                             & 15.2 & -0.296 & 1.04   & -3.15  & 0.393   & -0.893 \\
\hline
\hline
\end{tabular}
\caption{The same as Table~\ref{xsec8tabledyn} but at $\sqrt s=13$~TeV.}
\label{xsec13tabledyn}
\end{table*}

We now consider varying the Higgs boson width and couplings such that the cross section under the on-shell peak remains
constant.  The Higgs cross section at 8 TeV has the following form
for the case of the full set of cuts,
\begin{eqnarray}
\sigma^{H+I}_{\rm{full}}
 = 5.04+0.0395\left(\frac{\Gamma_H}{\Gamma_H^{\rm{SM}}}\right) \biggl(1-6.41\sqrt{\frac{\Gamma_H^{\rm{SM}}}{\Gamma_H}} \, \biggr)\, \rm{fb}
\label{eq:xs_hifull}
\end{eqnarray}
Note that, unlike in the $ZZ$ case, the constants appearing in front of the $\Gamma_H$-dependent terms in 
Eq.~(\ref{eq:xs_hifull}) cannot simply be inferred from the results given in Table~\ref{xsec8tabledyn}.  This is due to the
fact that $M_T$ is only a proxy for the actual quantity that appears in the Higgs boson propagator, $s_H$.
However the functional form of the equation is the same and so we have obtained Eq.~(\ref{eq:xs_hifull})
by fitting the cross section as a function of $\Gamma_H$.
Turning to the basic set of cuts the cross section can be parametrized by, 
\begin{eqnarray}
\sigma^{H+I}_{\rm{basic}}
 = 6.79+0.351\left(\frac{\Gamma_H}{\Gamma_H^{\rm{SM}}}\right) \biggl(1-3.33\sqrt{\frac{\Gamma_H^{\rm{SM}}}{\Gamma_H}} \, \biggr)\, \rm{fb}
\label{eq:xs_hibas}
\end{eqnarray}
The impact of the $m_{\ell\ell}$ and $\phi_{\ell\ell}$ cuts is clear from the above equations.  The application 
of these cuts leaves around 75\% of the resonant (NWA) cross section, compared to only 20\% and 10\% of the 
interference and off-shell cross sections respectively. This should be taken into consideration when designing future analyses
to be maximally sensitive to the rescaling parameter $\xi$.
In addition it is clear that the interference plays a very important role in these analyses.  For values of $\xi$ close to the Standard
Model it is the dominant effect and the expected number of Higgs events is reduced.  For very large $\xi$ the linear term is most
important and the net effect is an increase in the number of Higgs events expected.  For the basic set of cuts, 
Eq.~(\ref{eq:xs_hibas}) shows that this cross-over occurs for $\Gamma_H = 10 \, \Gamma_H^{\rm SM}$.

\section{Bounding the Higgs width using ATLAS $WW$ data} 
\label{sec:ATLAS}

In this section we use the ATLAS $H \to WW$ analysis of ref.~\cite{ATLAS:2013wla} to constrain the Higgs  
width. This channel is significantly more challenging than the equivalent 
analysis for $H \to ZZ$.  The existing experimental analyses are tailored to 
a search for on-shell Higgs boson events and only limited information is available regarding
the region at high transverse mass which is most sensitive to the Higgs boson width.
Therefore our results should only be viewed as indicative of those that may be obtained 
using a dedicated experimental analysis, which we aim to motivate with this study.

For simplicity we shall use only the $20\,{\rm fb}^{-1}$ of data taken at $8$~TeV and concentrate on
the $e\mu + \mu e$ channel, corresponding to the cuts applied in the previous section.
The ATLAS collaboration presents results for the expected  and observed number of events, including statistical uncertainties,
with the fiducial  cuts applied sequentially.   For the $N_{\rm jet}=0$ case in which we are interested,
the search for the Higgs boson is then performed in the transverse mass window, $0.75 \, m_H < m_T < m_H$. 
In this fiducial region ATLAS presents a full uncertainty analysis, accounting for a range of systematic effects.
For our analysis we wish to be sensitive to off-shell 
Higgs events so we are primarily interested in the data outside this $m_T$ window.  To proceed we must therefore
estimate the systematic uncertainty in the wider data set. 

We first consider the uncertainty on the background estimate.  We expect that a large uncertainty in the $N_{\rm jet}=0$ bin
should come from the normalization of the $WW$ background.   In their paper ATLAS present a total uncertainty on this quantity
that is derived in a control region which  is similar to our basic set of cuts.    We adopt this uncertainty, $7.4$\% as one
measure of the systematic uncertainty on the expected background ($\delta_B$), ignoring differences in uncertainties
associated with subdominant backgrounds and other systematic effects.  To account for this oversimplification we also consider the
more conservative choices $\delta_B=10\%$ and $\delta_B=12\%$.  The systematic uncertainty on the signal is dominated by 
the theory uncertainty and can be inferred from the detailed discussion in ref.~\cite{ATLAS:2013wla}, $\delta_S=21$\%.

Under the assumption $\delta_B=7.4\%$ we can thus estimate the absolute systematic uncertainty,
\begin{eqnarray}
N_{{\rm{full}}}^B({\rm{exp}}) = 1240 \pm 92 \,, \qquad  N_{{\rm{basic}}}^B({\rm{exp}}) = 5490 \pm 406  \nonumber
\end{eqnarray}
where we have read off the expected number of events under the full and basic set of cuts from Table 8 of ref.~\cite{ATLAS:2013wla}.
For the expected number of signal events we simply rescale our Eqs.~(\ref{eq:xs_hifull}) and~(\ref{eq:xs_hibas}) so that the result matches the
ATLAS expectation where, for consistency, the effect of the interference is neglected.  We thus find,
\beq
N^{H+I}_{{\rm{full}}}({\rm{exp}})=118.1+0.925\left(\frac{\Gamma_H}{\Gamma_H^{\rm{SM}}}\right) \biggl(1-6.41\sqrt{\frac{\Gamma_H^{\rm{SM}}}{\Gamma_H}} \, \biggr)
\nonumber
\eeq
for the full set of cuts and, for the basic cuts,
\beq 
N^{H+I}_{{\rm{basic}}}({\rm{exp}})=148.3+7.67\left(\frac{\Gamma_H}{\Gamma_H^{\rm{SM}}}\right) \biggl(1-3.33\sqrt{\frac{\Gamma_H^{\rm{SM}}}{\Gamma_H}} \, \biggr)
\nonumber
\eeq

\begin{center} 
\begin{figure*}
\includegraphics[scale=0.5,angle=0]{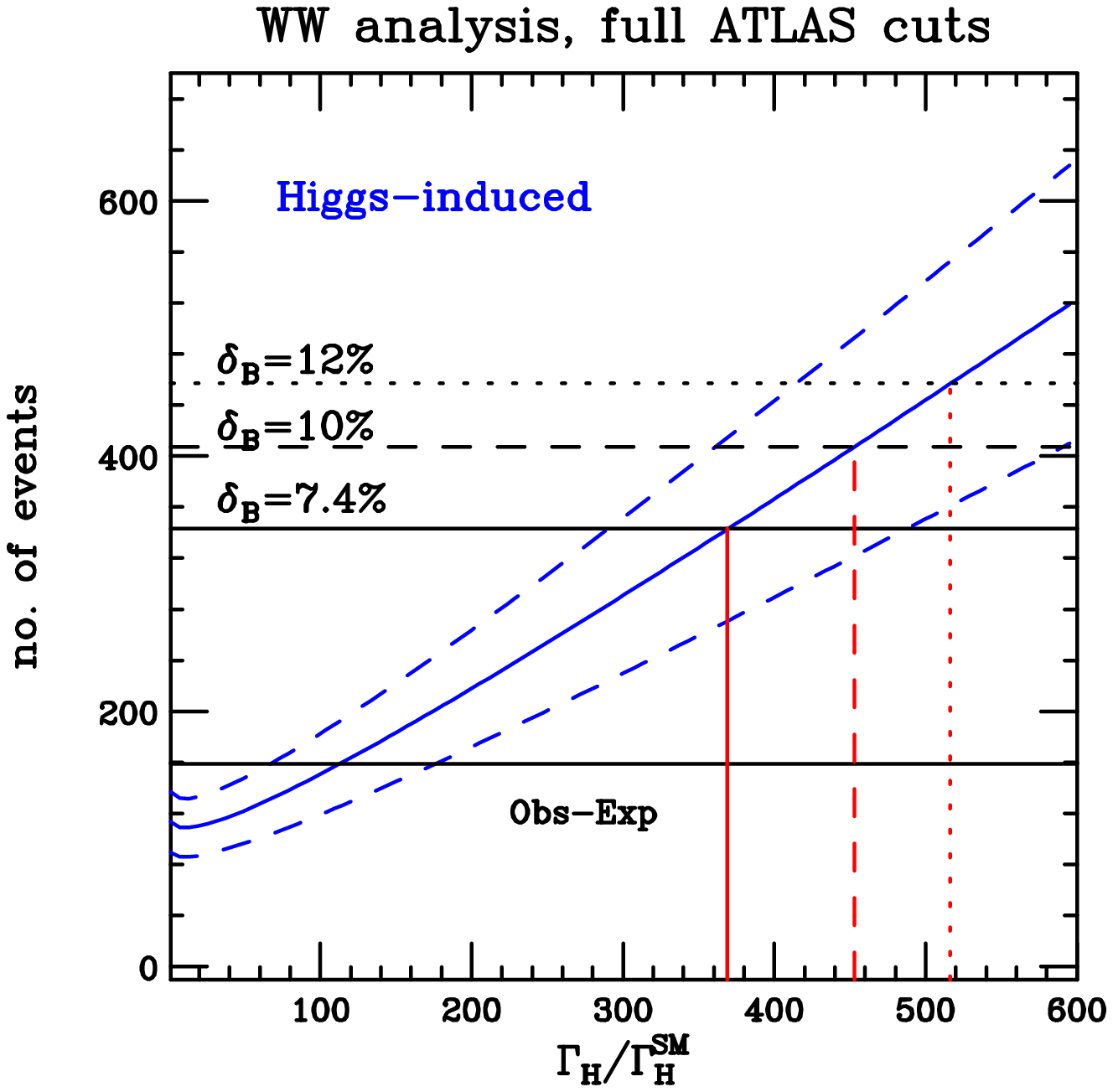} 
\includegraphics[scale=0.5,angle=0]{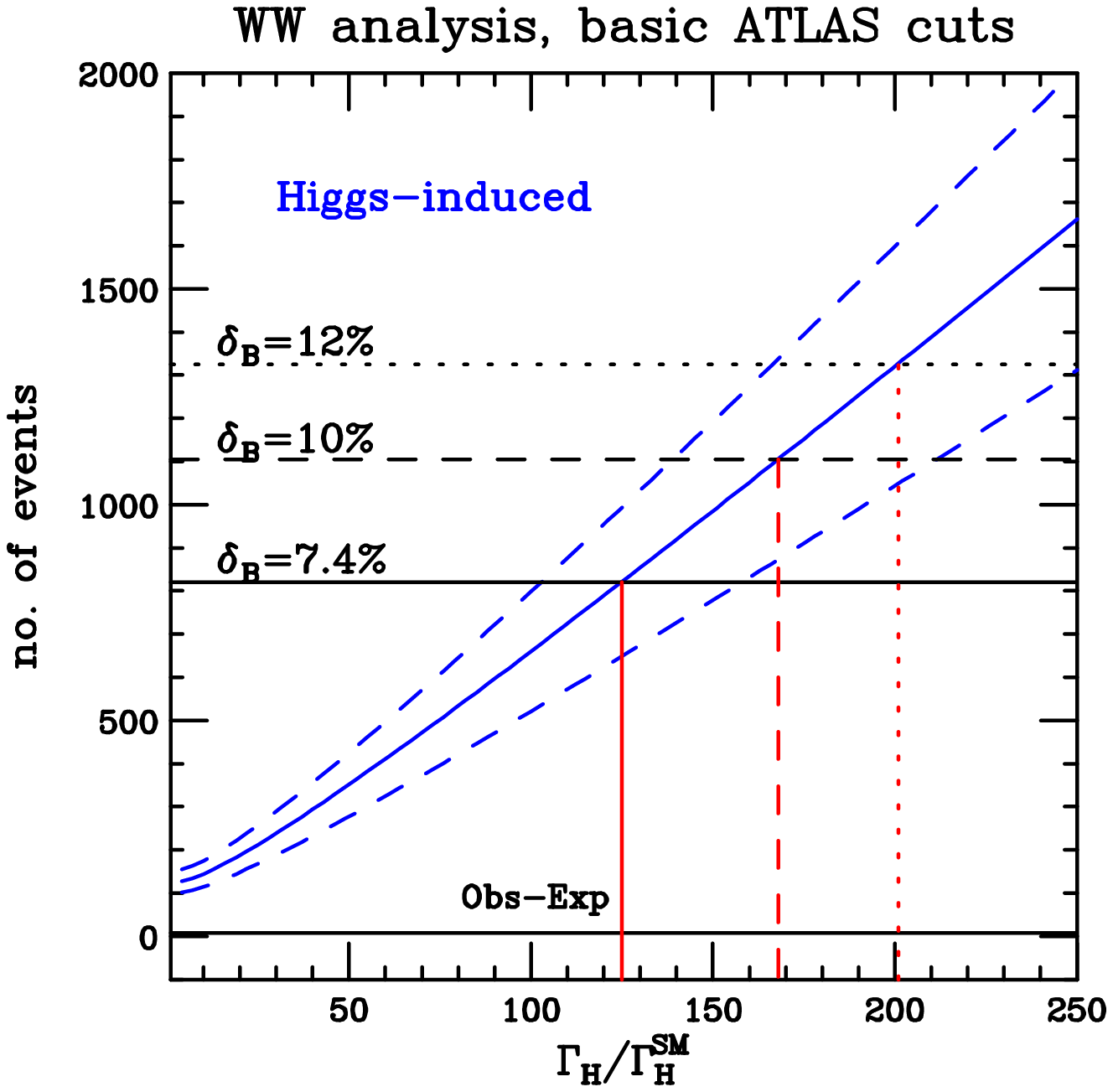} 
\caption{Limits on the Higgs width obtained using the results reported by ATLAS. The solid line represents the limit obtained using
an estimate of the systematic uncertainty obtained from the results presented in ref.~\cite{ATLAS:2013wla} ($\delta_B=7.4$\%).  The dashed and dotted lines represent
limits obtained using the more conservative choices $\delta_B=10$\% (dashed) and $\delta_B=12$\% (dotted). 
}
\label{hwidthWW}
\end{figure*}
\end{center}
These expectations can then be used to constrain the width of the Higgs boson given that $1399$ (full cuts) and $5497$ (basic cuts) events were observed
in the data.
Our results are summarized in Fig.~\ref{hwidthWW}, in which we also illustrate the sensitivity to the experimental systematics
by plotting results for $\delta_B=10\%$ and $\delta_B=12\%$.  Note that we have not considered the statistical uncertainty since it is small enough
that it has a negligible effect on the final results.
We can estimate the theoretical uncertainty by simply considering the uncertainty on the signal expectation discussed above. 
Using $\delta_B=7.4$\% and a $2\sigma$ excursion on the expected number of events we find, using the full cuts, 
\begin{eqnarray}
\Gamma_{{H}} \;  < \; 365^{+118}_{-79} \;\Gamma^{{\rm SM}}_H
\end{eqnarray}
at 95\% confidence.
From the discussion of the previous section we expect the limits obtained using the basic cuts to be much more sensitive to the width.
Indeed we find, 
\begin{eqnarray}
\Gamma_{{H}} \;  < \; 125^{+23}_{-22} \;\Gamma_{H}^{\rm SM} 
\end{eqnarray}
although this is also partly driven by the fact that the observed number of events is very close to that expected from background alone.
The limit obtained above is around a factor of three weaker than that obtained using a similar analysis 
of $ZZ$ data~\cite{Campbell:2013una}. 

In order to increase the sensitivity to a rescaling of the Higgs boson width it is crucial to focus on the events at large
transverse mass.   With that in mind we now consider an analysis that uses the basic ATLAS selection cuts defined above
but adds a simple transverse mass cut,  $M_T > 300$~GeV.
Under these cuts the Higgs-induced cross section is, 
\begin{eqnarray}
\sigma^{H+I}_{{M_T}}
= 0.004+0.241\left(\frac{\Gamma_H}{\Gamma_H^{\rm{SM}}}\right) \biggl(1-2.32\sqrt{\frac{\Gamma_H^{\rm{SM}}}{\Gamma_H}} \, \biggr)\, \rm{fb}
\label{eq:xs_hifull300}
\end{eqnarray}

\begin{center} 
\begin{figure*}
\includegraphics[scale=0.5,angle=0]{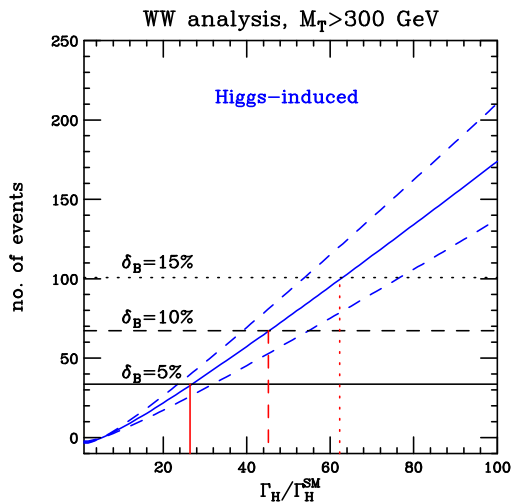} 
\caption{Estimated limits on the Higgs width obtained using a cut on the transverse mass, $M_T > 300$~GeV. The limits are computed
using an estimate of the expected background cross section, as described in the text, and for 
different projected experimental uncertainties. The number of signal events is also estimated from the ATLAS  20 fb$^{-1}$ 
expectation, with the corresponding theoretical uncertainty represented by the dashed curves.
}
\label{fig:highMT}
\end{figure*}
\end{center}

To obtain the expected number of signal events in the high $M_T$ region we simply rescale this cross section by the same
factor that is necessary to obtain the ATLAS expectation in the whole $M_T$ range, i.e. the factor that was used
for the basic cuts previously. 
In order to estimate the expected number of background events we compute the fraction of the continuum $WW$ background
process that survives the $M_T > 300$~GeV cut at leading order.  This fraction is around $6$\% and leads to an estimate
of 336 background events.  Rather than extrapolating our assumed background uncertainty from the previous analysis, we simply present estimates
of the expected limits that would be obtained with $\delta_B=5,10,15\%$.  We maintain the same theoretical uncertainty (21\%) on the 
Higgs signal. 
Our results are summarized in Fig.~\ref{fig:highMT}. 
As expected the large $M_T$ region is more sensitive to the Higgs width rescaling.  For instance, the expected 95\% confidence limit for
$\delta_B=10$\% is,  
\beq
\Gamma_H < 45^{+9}_{-7} \; \Gamma^{{\rm SM}}_H
\eeq
In comparison, the result from using a similar analysis in the $ZZ$ channel with $m_{4\ell} > 300$ GeV is around $25 \,\Gamma_H^{\rm{SM}}$~\cite{Caola:2013yja,Campbell:2013una}. 
These results suggest that, if the total experimental uncertainty can be constrained to below 10\%, then 
the $WW$ results can be complementary to, and even competitive with, those found in the $ZZ$ mode. 
Finally, we note that it is possible that the limits in the $WW$ channel could be further improved by including the $ee$ and $\mu\mu$ results. 
However, the analysis is already dominated by systematic uncertainties, so that limits obtained by including this data should not be significantly 
better than those suggested here.

\section{Conclusions}
\label{sec:conc}
In this paper we have motivated the use of existing LHC data to bound the Higgs width, using off-shell 
Higgs events in the $WW$ final state. The
analysis proceeds in a similar fashion to the $ZZ$ channel~\cite{Caola:2013yja,Campbell:2013una}.
The essential idea is that the cross section depends on the width of the Higgs boson in different ways near the peak
region and away from it.
This technique is optimal in instances in which one can directly measure the four-momentum squared of the Higgs boson, $s_H$,
thus allowing for a clean separation of the two regions of phase space.
In decays such as $H \to WW \to 2\ell2\nu$ this is impossible due to the presence of neutrinos that are identified only
as missing transverse momentum. However in these decays 
the transverse mass $M_T$ acts as an appropriate proxy for  $\sqrt{s_H}$.
The narrow peak in $\sqrt{s_H}$ is transformed into a broad excess in $M_T$, but 
the kinematics of the decay result in an edge at $M_T = m_H$. As a result there are very few on-shell events in the $M_T > m_H$ region. 

Using 20 fb$^{-1}$ of 8 TeV ATLAS $e\mu+\mu e$ data we investigated the potential for constraints on the Higgs width now and in the future. 
Since the existing analysis is 
dedicated to the search for, and measurement of, on-shell Higgs events, the current cuts are not ideal for our purpose. 
Nevertheless we estimate a bound on the Higgs width at approximately 125 times the SM value.
We illustrated how these limits 
could be significantly improved by focussing on the region of high transverse mass. 
If an experimental precision corresponding to a background uncertainty of $\delta_B \lsim 10\%$ could be achieved then the existing data 
may be able to bound the Higgs width at the level of $(25-50) \, \Gamma_H^{\rm SM}$. 
For this reason we believe that a full experimental analysis, focussing particularly on the high-$M_T$ region of the $N_{\rm jet}=0$ bin of the $WW$ channel,
is extremely well motivated.

\section*{Acknowledgements}
\noindent
The research of RKE and JMC is supported by the US DOE under contract DE-AC02-07CH11359.

\bibliography{WW_prd}

\end{document}